# Optical Switching of Moiré Chern Ferromagnet

Xiangbin Cai[1,2#], Haiyang Pan[1,2#], Yuzhu Wang[2#], Abdullah Rasmita[1,2], Shunshun Yang[1], Yan Zhao[1,2], Wei Wang[3], Ruihuan Duan[4], Ruihua He[1], Kenji Watanabe[5], Takashi Taniguchi[6], Zheng Liu[4], Jesús Zúñiga Pérez[1,7], Bo Yang[2]*, Weibo Gao[1,2,8,9]*

[1]School of Electrical & Electronic Engineering, Nanyang Technological University, Singapore 639798, Singapore

[2]Division of Physics and Applied Physics, School of Physical and Mathematical Sciences, Nanyang Technological University, Singapore 637371, Singapore

[3]Key Laboratory of Flexible Electronics (KLoFE) & Institute of Advanced Materials (IAM), School of Flexible Electronics (Future Technologies), Nanjing Tech University, Nanjing 211816, China

[4]School of Materials Science and Engineering, Nanyang Technological University, Singapore 639798, Singapore

[5]Research Center for Electronic and Optical Materials, National Institute for Materials Science, 1-1 Namiki, Tsukuba 305-0044, Japan

[6]Research Center for Materials Nanoarchitectonics, National Institute for Materials Science, 1-1 Namiki, Tsukuba 305-0044, Japan

[7]Majulab International Joint Research Unit UMI 3654, CNRS, Université Côte d'Azûr, Sorbonne Université, National University of Singapore, Nanyang Technological University, Singapore

[8]Center for Quantum Technologies, Nanyang Technological University, Singapore 637371, Singapore

[9]Quantum Science and Engineering Center (QSec), Nanyang Technological University, Singapore 639798, Singapore

[#]These authors contributed to this work.

*Correspondence to B.Y. (yang.bo@ntu.edu.sg) and W.G. (wbgao@ntu.edu.sg).

**Abstract**

Optical manipulation of quantum matter offers a non-contact, high-precision and fast control. Fractional Chern ferromagnet states in moiré superlattices are promising for topological quantum computing, but an effective optical control protocol has remained elusive. Here, we demonstrate robust optical switching of integer and fractional Chern ferromagnets in twisted $MoTe_2$ bilayers using circularly polarized light. Highly efficient optical manipulation of spin orientations in the topological ferromagnet regime is realized at zero field using a pump light power as low as 28 nW·µm$^{-2}$. Utilizing this optically induced transition, we also demonstrate magnetic bistate cycling and spatially resolved writing of ferromagnetic domain walls. This work establishes a reliable and efficient optical control scheme for moiré Chern ferromagnets, paving the way for dissipationless spintronics and quantized Chern junction devices.

**Main Text**

Control of magnetism has long been a goal of fundamental and practical importance throughout science and engineering. The last decade has witnessed the rise of two-dimensional (2D) magnetic materials [1-5], which feature drastic quantum confinement effects in the out-of-plane direction and thus uncover a plethora of exotic quantum phases. Extensive efforts have been made to manipulate 2D magnetic states, including magnetic phase transitions by applying electrostatic doping and electric fields [6-9], tuning interlayer stacking order [10,11], exploiting mechanical strain [12-15] as well as optical excitation [16-19]. On the other hand, all-optical control of magnetism has attracted growing attention because of its non-contact nature, high spatial precision, fast responsivity and selective-interaction behavior by fine-tuned laser excitation wavelength and polarization [20-25]. Notably, high-efficiency non-volatile optical training of ferromagnetic domain formation has been demonstrated in layered $Fe_3GeTe_2$ using a low power density of around 20 $\mu W \cdot \mu m^{-2}$, which utilizes optically injected spin-polarized electrons as initial magnetic seeds [26]. Optical control of atomically thin magnets can lead to next-generation compact and ultrafast coherent spin-control technologies.

Beyond conventional magnetism in layered materials originating from magnetic elements, there have been exciting works on interaction-driven ferromagnetism (FM) in moiré superlattices [27-29], formed by stacking 2D non-magnetic layers with a twist angle or lattice mismatch. The recent discovery of fractional quantum anomalous Hall effects in twisted 1L/1L $MoTe_2$ bilayers and few-layer rhombohedral graphene/h-BN superlattices has further ignited the community with emergent moiré Chern ferromagnets [30-34], where the spontaneously spin-polarized ground states support topological phases exhibiting charge fractionalization. Their spin-valley-locked flat Chern bands suggest promising topological phase control by manipulating the ferromagnetic order [35-38]. In particular, the highly tunable quantum geometry effect [39] and strong electronic correlations [40,41] in the flat Chern bands of twisted $MoTe_2$ bilayers may potentially host topological superconductivity [42,43] and non-Abelian anyons [44], which are critical building blocks for fault-tolerant topological quantum computation. However, the optical switching and topological manipulation of moiré Chern ferromagnets remains elusive, an essential step towards on-demand design of quantum matter and topological devices [45].

In this report, we demonstrate a robust optical control strategy of moiré Chern ferromagnetic states in rhombohedral (R) small-angle twisted $MoTe_2$ bilayers using circularly polarized light (CPL).

The spin-valley locking and spontaneous time-reversal-symmetry (TRS) breaking of moiré Chern bands in twisted MoTe2 enables zero-field optical initialization of the FM order when cooled down below the Cuire temperature, where the FM orientation is dependent on the helicity of CPL. Highly efficient optical switching of correlated FM is demonstrated across the phase space of filling factor and electric field, using an ultralow pump power of 28 nW·μm$^{-2}$. The magnetization switching efficiency substantially depends on the incident photon energy and polarization, and reaches close to 100% for a CPL in resonance with the absorption feature of hole-doped twisted MoTe2 bilayers. Additionally, reiterated bidirectional switching cycles of FM states confirm the robustness of this optically induced magnetic phase transition. High spatial-resolution patterning of FM domain walls has also been demonstrated for programmable Chern junction [46] and topological memory [38] applications. Our results establish a highly efficient non-volatile optical control protocol for moiré Chern ferromagnets, and open a new avenue for exploring magneto-optical interactions in topological quantum matter.

**Zero-Field Initialization of Moiré Chern Ferromagnet**

The 3.50° (device #1) and 3.65° (device #2) twisted 1L/1L MoTe2 bilayers are encapsulated within the graphite/h-BN dual-gate structure for independent electrical control of filling factor *v* (charge carrier number per moiré unit cell) and electric field $D/\varepsilon_0$ in the superlattice, which are calculated using the parallel plate capacitor model. As shown in **Extended Data Figure 1**, a pump-and-probe measurement scheme was adopted for the optical FM switching experiment, where a CPL pump beam is aligned with the reflective magnetic circular dichroism (RMCD) probe. The sign and amplitude of the RMCD signal accurately track the spin orientation and the magnetization strength in samples, respectively [29,30,47]. Detailed device fabrication processes and optical measurement workflows can be found in the Methods section. Most results in the main text are obtained from device #1, which exhibits a moiré density of 3.45 x 10$^{12}$ cm$^{-2}$ with a moiré size of 5.78 nm.

The long-wavelength moiré periodicity in small-angle-twisted 1L/1L MoTe2 superlattice enables a continuum description of its low-energy electronic structure. The valence band edge of twisted MoTe2 bilayer resides at the corners of the moiré Brillouin zone, i.e., *K* and *K'* points, whose continuum Hamiltonians are related by TRS [35,41]. As shown in **Figure 1 (a)** and **(b)**, the exchange interactions in the nearly flat topmost valence bands spontaneously produce ferromagnetism [30,48-

[50] at $v = -1$, and at fractional fillings such as $v = -2/3$, stabilized fractional Chern insulating states arising from significant electronic correlation. The strong spin-orbit coupling (SOC) in MoTe$_2$ locks the spin to the valley degree of freedom, so σ+ CPL couples with the *K*-valley while σ- CPL couples with *K'*-valley [51,52]. Specifically, as the σ- case illustrated in **Figure 1 (b)**, if the sample at $v = -1$ has the gate-induced spin-up holes occupying the *K* valley, the σ- CPL pumping in resonance with the absorption feature of the system will excite electron-hole pairs into the *K'* valley. The dispersive conduction band minima at *K/K'* valleys exhibit weak spin-splitting and may allow inter-valley electron scattering [53], while the inter-valley hole transfer between moiré Chern bands is suppressed by the opposite spin and Chern numbers. As a result, the optically injected spin-polarized holes form a long-lived valley polarization in the *K'*-valley, even after the pump light is turned off. Once this optically generated valley/spin imbalance exceeds a critical density, the exchange interaction produces an effective Zeeman-like field that overcomes the existing valley splitting and reverses the FM order to the *K'*/spin-down state as shown in the lower panel of **Figure 1 (b)**. The proposed mechanism of the optical FM switching in the σ- CPL pumping case is further illustrated step by step in **Extended Data Figure 2**, and the σ+ CPL pumping case should be similar.

Utilizing this mechanism to nucleate FM domain seeds of designated spin orientations, zero-field all-optical initialization of the moiré Chern ferromagnet has been demonstrated in **Figure 1 (c)**, in contrast to the conventional magnetic field training. Below the Curie temperature of around 12.5 K at $v = -1$, the σ- CPL pumping effectively induces the formation of a uniform magnetic domain in the spin-down state (manifested as increasingly positive RMCD), while the σ+ CPL induces a spin-up polarized state (manifested as increasingly negative RMCD). After the σ- initialization to the cryostat base temperature (1.5 K), the correlated FM states in twisted MoTe$_2$ superlattice are verified by photoluminescence (PL) mapping across the $v$ - $D/\varepsilon_0$ phase space as shown in **Figure 1 (d)**, respectively. The PL map shows lower intensity at constant carrier density of $v = -1$ and -2/3 within certain electric field strength ranges. The reason is that the formation of integer and fractional Chern insulators, respectively, opens a bulk gap that reduces the free carrier density available for the trion emission [30,54]. The zero-field RMCD mapping across the $v$ - $D/\varepsilon_0$ phase space under 28 nW·μm$^{-2}$ σ- and σ+ CPL pumping are shown in **Figure 1 (e)** and **(f)**, respectively. Both maps exhibit correlation-promoted FM at $v = -1$ and -2/3. The typical wing shape of correlated FM at $v = -1$ and -2/3 reflects the electric-field-induced phase transition from Chern insulators to topological trivial states, together with the breaking of layer degeneracy. Ramping the magnetic field at $v = -1$ and $-2/3$ yields the characteristic magnetic hysteresis loops showing coercive field strengths (i.e., the external magnetic field strength when the magnetization vanishes) of around 51 and 10 mT, respectively, as shown in **Extended Data Figure 3**, where detailed characterization of the correlated FM states in device #1 can be found.

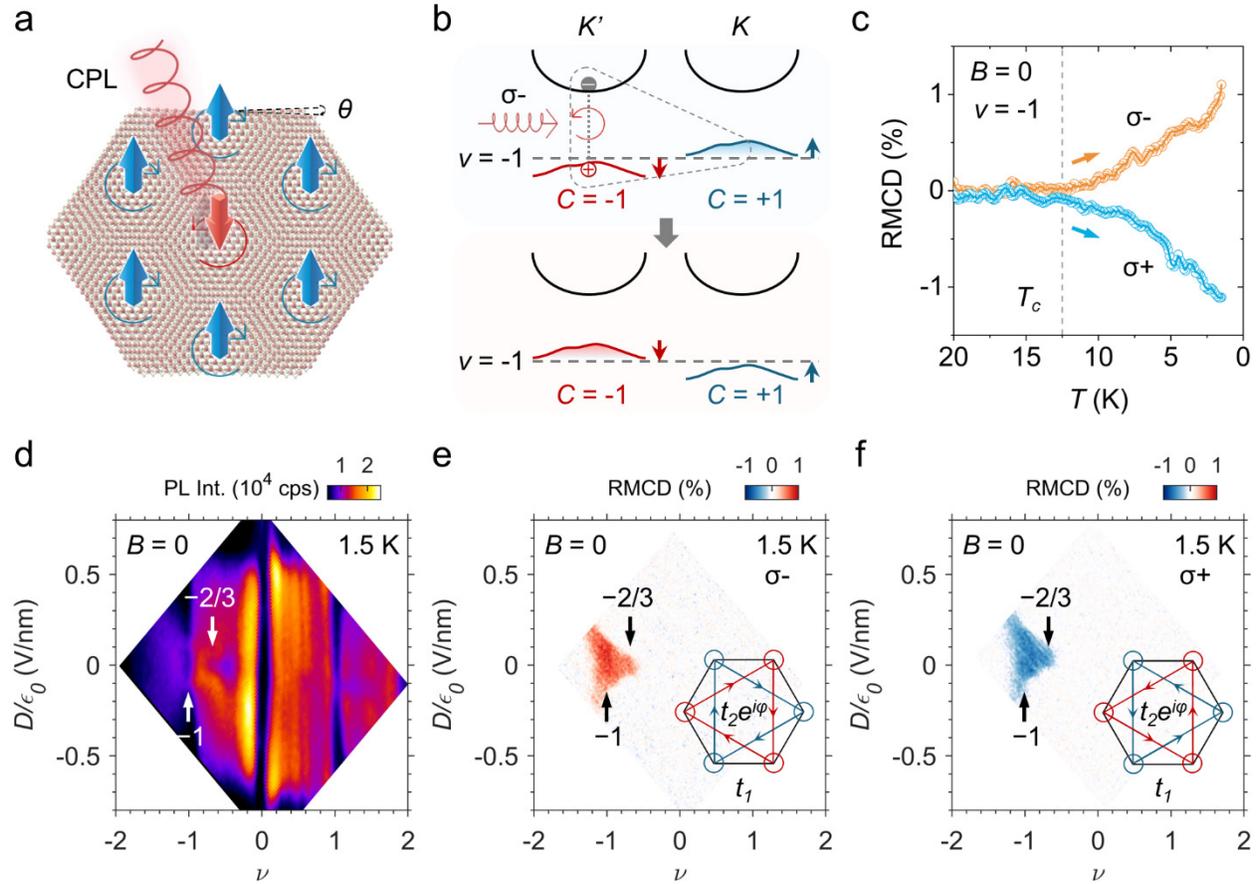

**Figure 1: Zero-field initialization of moiré Chern ferromagnet. (a)** Schematic representation illustrating local spin flip in the moiré Chern ferromagnet induced by the circularly polarized light (CPL). The arrows represent spin orientations in each moiré unit cell at filling factor $v = -1$. **(b)** Valley-resolved band structure of the moiré Chern ferromagnet at $v = -1$. The CPL reverses the spin population between valleys. The spin orientations (represented by arrows) and the Berry curvature signs (indicated by contrasting colors) of the topmost valence bands are locked to the valley ($K$ and $K'$) degree of freedom, and couple to opposite helicities of CPL. **(c)** Temperature-

dependence of the reflective magnetic circular dichroism (RMCD) signal of 3.50°-twisted MoTe$_2$ bilayer (device #1) at $v = -1$, pumped with 28 nW·µm$^{-2}$ continuous-wave (CW) CPL before each measurement (see Methods for experimental details). **(d)** Photoluminescence (PL) count rate map of device #1 with varying charge filling factor $v$ and out-of-plane electric field $D/\varepsilon_0$, showing correlated phases as dim vertical lines. RMCD contrast maps of device #1 with varying $v$ and $D/\varepsilon_0$ under 28 nW·µm$^{-2}$ **(e)** σ- and **(f)** σ+ CPL pumping, respectively, at zero field. The insets illustrate that two sets of degenerate energy minima (red and blue circles) at moiré high-symmetry points form a honeycomb superlattice, where chiral complex hopping between next-nearest-neighbor sites accounts for the moiré Chern ferromagnet [30,55].

**Optical Switching of Correlated Ferromagnetism**

The optical FM switching mechanism enables direct reversal of the magnetization state in moiré Chern ferromagnets below the Curie temperature, where the switching effect is determined by the CPL pump power, i.e., the CPL-induced spin-polarized charge density. As illustrated in **Figure 2 (a)**, a pump-probe sequence is applied to examine the FM switching effect under varying pump powers and electronic phases. Before the pump is turned on within each pump-probe cycle, a reset process is applied, which includes (1) gates ramp to designated $v$ and $D/\varepsilon_0$ conditions; (2) an initializing pump pulse (i.e. 28 nW·µm$^{-2}$ σ- CPL irradiation for 0.5 s) to induce the spin-down orientation in the sample; (3) a 0.5 s dark time (i.e. without any optical excitation) for the sample to relax from thermal and electrical perturbation. Then, a σ+ CPL of tunable power is turned on for 0.5 s to exert spin angular momentum transfer onto the sample, and another 0.5 s dark time to rest the sample is respected before the weak probe beam is turned on to detect the magnetic state. Additional details of measurement workflows can be found in the Methods section.

In **Figure 2 (b)**, without any σ+ CPL pumping, the RMCD signal peaks locally at $v = -1, -4/5$ and $-2/3$ due to the interaction-promoted FM at zero magnetic field. Upon a 7 nW·µm$^{-2}$ σ+ CPL pump excitation, the spin-down orientation (positive RMCD) of the system at those correlated FM states flips upwards (negative RMCD), while the FM of metallic phases at other fillings is suppressed. With increasing pump power, the switched FM reaches saturation of inversed RMCD at 28 nW·µm$^{-2}$, three orders smaller than the reported optical training of itinerant two-dimensional magnets [26]. The switchability of FM metallic phases presented here excludes the exciton-mediated exchange mechanism, which also requires continuous CPL excitation for the FM order [17]. The ultrahigh efficiency of direct optical switching of $v = -1$ and $-2/3$ moiré Chern ferromagnets is further verified against $D/\varepsilon_0$ in **Figure 2 (c)** and **(d)**, which exhibit consistent saturated FM flipping with 28 nW·μm$^{-2}$ CPL pumping. Although the FM flipping occurs with 7 nW·μm$^{-2}$ CPL pumping, the RMCD for $+ D/\varepsilon_0$ and $- D/\varepsilon_0$ after switching behaves asymmetrically, which might be due to the slight device asymmetry associated to unequal dielectric environments for the upper and lower MoTe$_2$ layers.

The optical switching of FM states in moiré Chern ferromagnets by CPL exhibits characteristics of a first-order phase transition, as demonstrated by the hysteresis loops shown in **Extended Data Figure 4 (a)** and **(b)**. The critical pump powers for completely flipping the magnetization at $v = -1$ and $-2/3$ are observed to be around 20 and 17 nW·μm$^{-2}$, respectively. The reproducibility of direct optical switching of moiré Chern ferromagnet is further confirmed in another device, i.e., 3.65°-twisted 1L/1L MoTe$_2$ bilayer (device #2) as shown in **Extended Data Figure 5**.

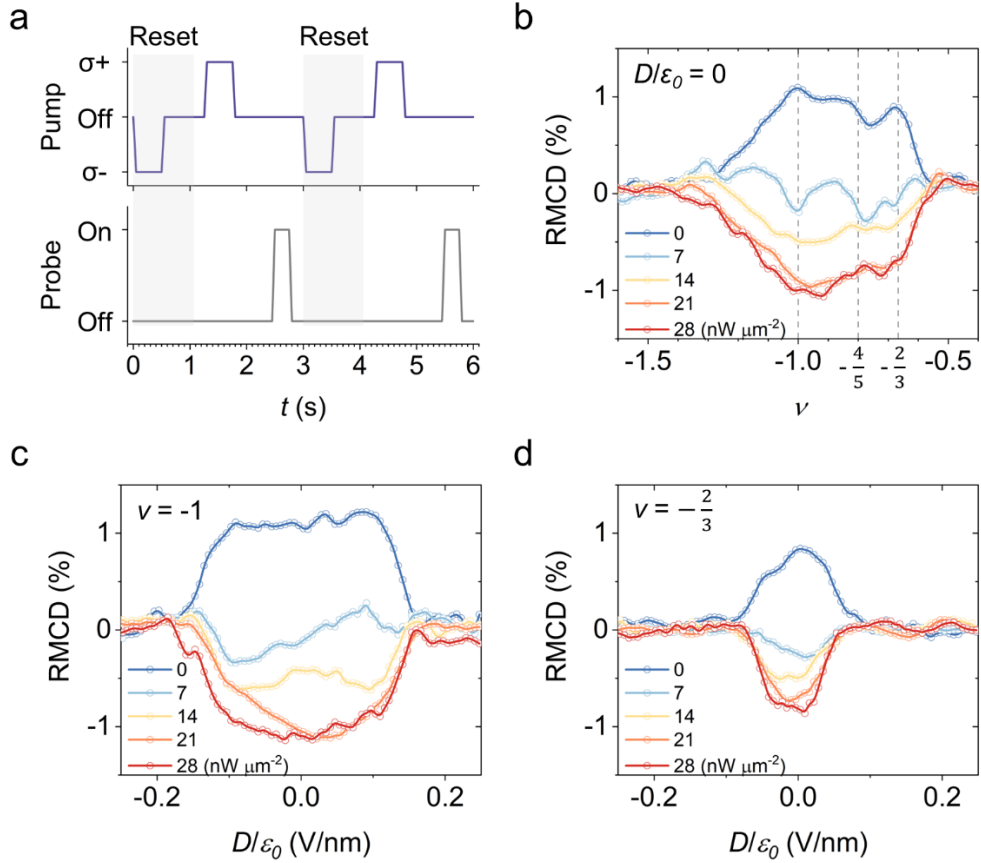

**Figure 2: Optical switching of correlated ferromagnetism. (a)** Measurement sequence for the optical switching of magnetic states in hole-doped twisted MoTe2. The magnetic field is kept equal to zero during the whole process. The "reset" stage includes gate ramping to achieve designated filling factor and electric field, zero-field initialization of sample's ferromagnetic state by 28 nW·μm$^{-2}$ σ- induction and a subsequent dark rest time, as described in the Methods. **(b)** Filling-factor-dependence of the optical switching of sample's FM states using varying σ+ pumping powers at zero electric field. Electric-field-dependence of the optical switching for **(c)** $v = -1$ and **(d)** $v = -2/3$ FM insulators using varying σ+ pumping powers in device #1, respectively. Symbols are measured values while solid lines are spline lines.

**Efficiency and Robustness of Optical FM Switching**

The efficiency of direct optical FM switching by CPL critically depends on the photon energy and polarization state of pump light. The switching efficiency is calculated as the percentage change of sample magnetization between the two states before and after pumping. A switching efficiency greater than 50% indicates that the sample magnetization flips its sign. Compared to the reflection contrast measurement as shown in **Extended Data Figure 6**, the pump photon energy range for >50% switching efficiency, as shown in **Figure 3 (a)**, matches the absorption energy of hole-doped twisted MoTe$_2$ bilayers. Furthermore, when the polarization state of pump light deviates from the σ+ circular polarization, the switching efficiency decreases monotonously, as shown in **Figure 3 (b)**. The σ- CPL pumping does not affect the FM order since the coupled spin-down moiré Chern band in the *K'* valley is already fully filled by gate-induced resident holes at $v = -1$. But when a linearly polarized light (LPL) pumping (±45° QWP angle in **Figure 3 (b)**) is applied, the $v = -1$ FM is suppressed, because the LPL can be regarded as a combination of equal σ+ and σ- CPLs, and the equally injected spin populations in both valleys disrupt the net magnetization, even when the LPL pumping is removed.

Both pump energy and polarization dependences of the FM switching efficiency align well with our proposed switching mechanism by optical spin pumping, highlighting the critical role of spin angular momentum transfer between optically excited and resident holes in spin-polarized Chern bands. The mechanism differs fundamentally from the previously reported optical FM switching by the inverse Faraday effect (IFE) [20,22], which typically requires a giant peak power density of $10^9$-$10^{12}$ nW·μm$^{-2}$. Since *K* and *K'* valleys are time-reversal partners, the optical FM switching is reversible as shown in **Figure 3 (c)**. The reiterated bidirectional switching of sample's FM state at $v = -1$ using 28 nW·μm$^{-2}$ alternating σ+/σ- pumping demonstrates the robustness of this optical switching scheme and the stability of switched FM states.

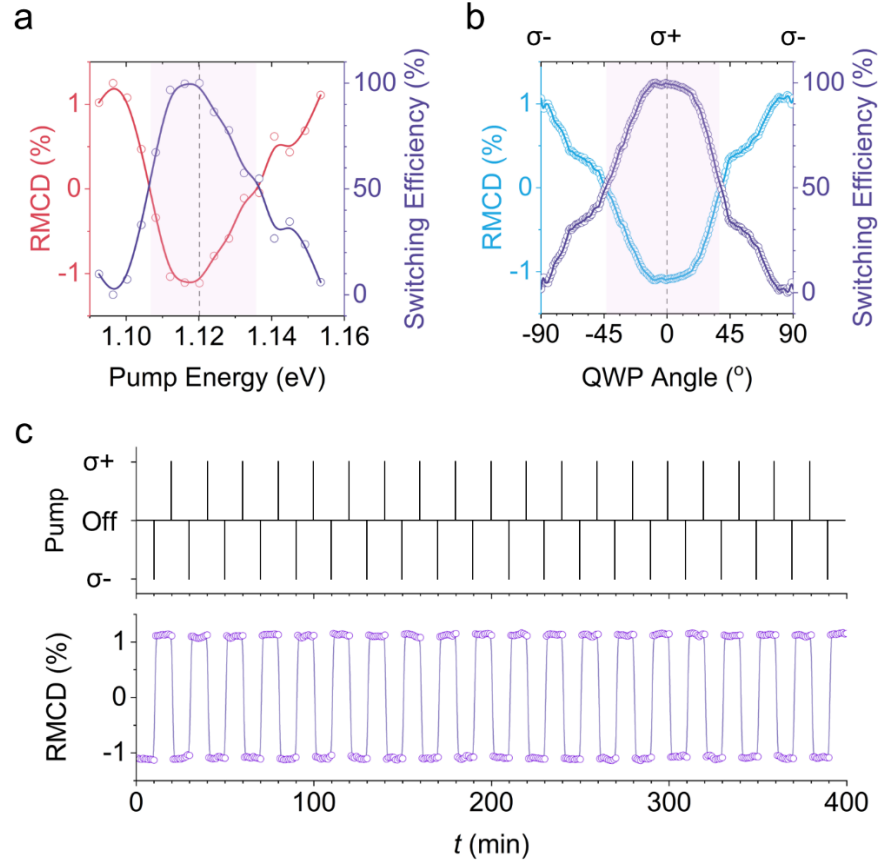

**Figure 3: Efficiency and robustness of optical FM switching. (a)** Photon energy dependence of optical switching of the FM state at $v = -1$ using 28 nW·µm$^{-2}$ σ+ pumping. The switching efficiency is defined as the percentage change of sample magnetization after pumping. The gray dashed line indicates the pump energy adopted for other optical switching measurements throughout the paper. The purple shaded range marks the pump energy window enabling >50% FM switching, where the sample magnetization flips its sign. **(b)** Pump polarization state dependence, determined by the quarter wave plate (QWP) angle, of optical FM switching at $v = -1$ using 28 nW·µm$^{-2}$ σ+ pumping. The gray dashed line indicates the QWP angle corresponding to a perfect σ+ configuration and adopted for other optical switching measurements throughout the paper. The purple shade marks the range of QWP angle enabling >50% FM switching, where the sample magnetization flips its sign. **(c)** Reiterated FM state switching at $v = -1$ using 28 nW·µm$^{-2}$ σ+/σ- pumping, demonstrating the robustness of our optical switching scheme. The magnetic field and electric field are kept equal to zero for all tests here.

**FM Domain Wall for Programmable Chern Junction**

After demonstrating the efficient and robust optical switching of correlated FM in twisted MoTe$_2$ bilayers, we exploited its high spatial precision for deterministic writing of FM domain walls and constructing arbitrary junctions with opposite magnetization as shown in **Figure 4**. The quantized anomalous Hall effect in a moiré Chern ferromagnet, including the twisted MoTe$_2$ superlattice, is characterized by a topological number (i.e. Chern number $C$) and the existence of a dissipationless chiral edge current, whose sign is determined by the magnetization direction [56-58], as illustrated in **Figure 4 (a)** and **(d)**. Since the topological number must change discontinuously at the domain wall separating spin-down and spin-up magnetic domains, a chiral edge current with predefined conductance quantum should appear at the domain wall [59,60]. In **Figure 4 (b)** and **(c)**, it can be seen that such FM domain walls, hosting topologically protected chiral edge modes, can be arbitrarily written in space using direct optical FM switching in a fractional Chern insulator, forming the basis of programmable Chern junctions [46] and topological domain wall devices [59-61] for quantum information technologies.

Although the direct measurement of quantized Hall resistance or dissipationless edge conduction before and after optical FM switching remains challenging due to electrical contact problems, we verified the topological nature of the optically written states using magnetic-field-dependent PL spectroscopy, as shown in **Extended Data Figure 7**. The extracted Streda slope is -1.21 ± 0.23 for the $v$ = -1 state, demonstrating that the topological gap remains intact after optically switching the FM orientation, which is reasonable given the rather weak pumping power density.

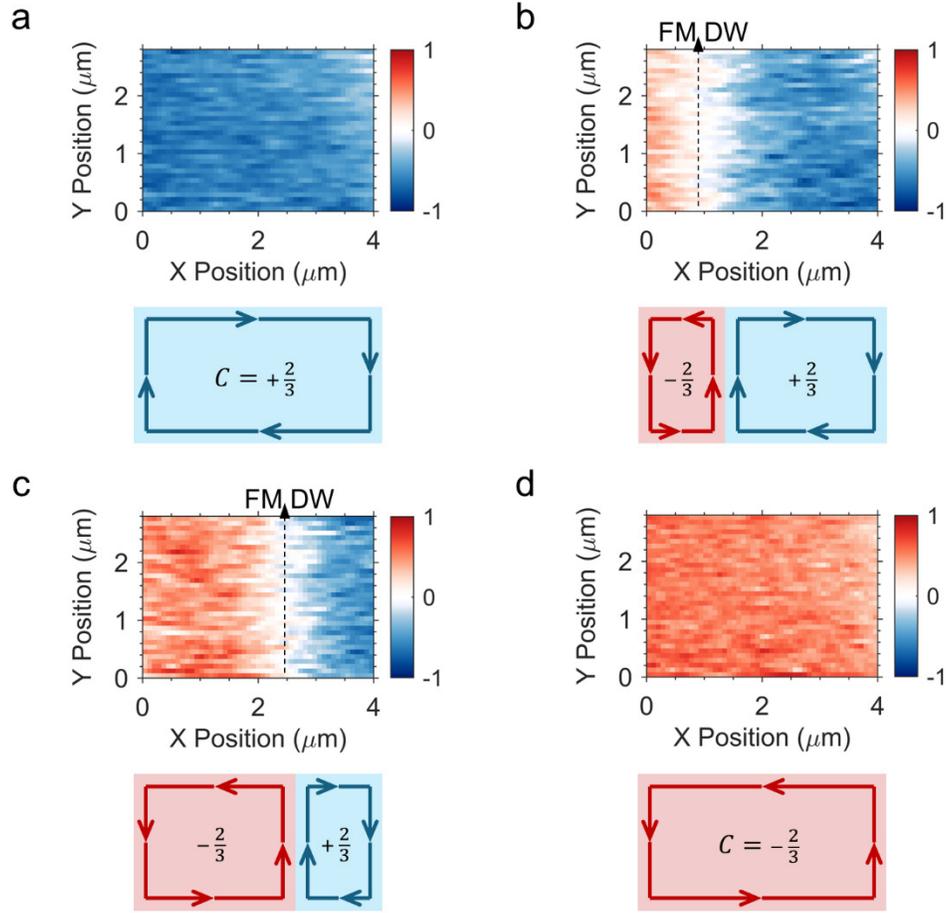

**Figure 4: FM domain wall for programmable Chern junction.** Spatial RMCD maps of the sample's FM states (upper panels) and the corresponding chiral edge current configurations (lower panels) at $v$ = -2/3. The evolution proceeds as follows: **(a)** full initialization of the sample into a spin-up FM state via 28 nW·µm$^{-2}$ σ+ CPL scanning; **(b)** partial reversal of magnetization near the left edge; **(c)** progressive move of the domain wall towards the right edge; **(d)** complete reversal across the entire channel region by column-by-column writing using 28 nW·µm$^{-2}$ σ- CPL. The FM domain wall is pushed from the left edge progressively towards the right edge and manipulates the chiral edge current configurations in the Chern junction. The magnetic field and electric field are kept equal to zero for all tests here.

**Conclusion and Outlook**

In summary, we have established an efficient and robust optical control protocol for moiré Chern ferromagnets using CPL excitation. The spin-valley locking and spontaneous TRS breaking of flat Chern bands in twisted MoTe$_2$ enables zero-field optical initialization of the ferromagnetic order, which is dependent on the helicity of pump light, when cooled down below the Curie temperature. Highly efficient optical manipulation of spin orientations in the topological ferromagnet regime is realized across the filling-factor and electric-field phase space, using a pump power density as low as 28 nW·µm$^{-2}$. The FM switching efficiency substantially depends on the incident photon energy and polarization state, and approaches close to 100% when a CPL is resonant with the absorption energy of the hole-doped system. Moreover, reiterated bidirectional switching of FM states proves the robustness and reliability of such an optically induced magnetic phase transition. High spatial-resolution construction of FM domain walls has also been demonstrated, opening the door for programmable Chern junctions and topological domain wall devices.

Looking ahead, we can explore the nature of topological chiral edge currents dictated by the two topological phases realized on either side of the domain wall. In particular, our approach can conveniently design two fractional topological phases with co-propagating and counter-propagating chiral edge currents carrying fractional charges interacting with each other at the domain wall. These interesting dynamics with edge reconstruction lead to new neutral and charge modes at the domain wall, that are difficult to realize in conventional quantum Hall systems, where only the edge between topological phase and vacuum is generally realized. They are important for understanding the stability of topological phases, braiding statistics of anyons and the robustness of quantum information processing.

## Methods

### Device Fabrication

Single crystalline 2H MoTe$_2$ was grown using the chemical vapor transport (CVT) method. The quality of synthesized flakes was checked and screened by the intensity and linewidth of neutral exciton PL in exfoliated monolayers before the parent crystal was selected for the dual-gate device fabrication. MoTe$_2$ monolayers, thin graphite layers (a few nm) and h-BN flakes (15-30 nm) were mechanically exfoliated onto silicon wafers with 285 nm SiO$_2$ layer, and identified by their optical contrast for clean and uniform regions. The bottom gate structure was first constructed by stacking BN/thin graphite on the SiO$_2$/Si wafer with pre-deposited 3/30 nm Cr/Au connecting electrodes. A grounding pin of thin graphite was further stacked onto the back gate surface with connection to another Cr/Au electrode. The contact-mode atomic force microscopy (AFM) was applied to clean the surface of bottom gate structure.

A sharp AFM tip was mounted onto a polydimethylsiloxane (PDMS) stamp in the transfer stage to cut the MoTe$_2$ monolayer flake. Subsequently, the twisted MoTe$_2$ bilayer was assembled by sequentially picking up monolayers with a desired twist angle using the BN/Polycarbonate (PC)/PDMS stamp by a fully motorized transfer stage. Stacking is performed without heating to reduce the interface degradation and slipping. After top BN encapsulation, the heterostructure was cleaned again by the contact-mode AFM to squeeze out residue and bubbles trapped in the interfaces, significantly increasing the moiré uniformity as reported elsewhere [30,31]. Finally, a thin graphite was placed over the twisted region and connected to a third Cr/Au electrode to serve as the top gate. The top and bottom BN thicknesses were confirmed by AFM measurements. All processes related to the air-sensitive MoTe$_2$ were conducted in the MBRAUN glovebox with the controlled concentrations of water and oxygen below 0.1 ppm (99.999% Ar cylinders as the gas source). To reduce the severe electrostatic charging in such a water-free environment, two DESCO Zero-Volt Ionizers were deployed besides the transfer stage. The fully encapsulated structure was rinsed in high-purity chloroform for 5 min to remove polymers before optical measurements.

### Optical FM Switching

Devices were mounted on a home-made chipset and inserted into the closed-loop helium cryostat

(attoDRY 2100), equipped with an attocube high-precision piezo stage, a 9 T superconducting magnet and supporting a base temperature of 1.5 K. The confocal pump-probe setup, as shown in the schematic of **Extended Data Figure 1**, utilizes an attocube apochromatic cryogenic objective of high numerical aperture (LT-APO-IR-0.81), which is non-magnetic to minimize the magnetic-field-induced drift. The dual gates were supplied by a Keithley 2636B dual-channel source meter. The pump beam was aligned with the focused probe beam using a near infrared (NIR) camera. The pump power density was kept at 28 nW·µm$^{-2}$ for zero-field induction as shown in **Figure 1 (c)** and most FM switching, unless specified otherwise.

After initializing the spin orientation of samples by the zero-field CPL induction, both magnetic field and circularly polarized pump light were turned off. For each optical FM switching and probing sequence, as diagramed in **Figure 2 (a)**, charge carrier density and electric field were first set by ramping gates. After 0.5 s of rest time, the pump light with the assigned helicity was turned on for 0.5 s. Another 0.5 s of rest time elapses before the probe light was turned on to detect the spin orientation in the RMCD scheme, where a 6 nW·µm$^{-2}$ narrow-linewidth (<1 MHz) laser from the continuous-wave optical parametric oscillator (C-Wave OPO by HÜBNER Photonics) was employed in resonance with the absorption feature of hole-doped twisted MoTe$_2$ bilayers. The probe intensity was modulated by a mechanical chopper at a frequency of $p$ = 937 Hz and the probe laser phase was modulated by 0.25$\lambda$ using a photo-elastic modulator (PEM-200 by Hinds Instruments) at a frequency of $f$ = 50 kHz. The reflected laser was collimated and focused onto an InGaAs avalanche photodiode detector (Thorlabs APD430C). The current output signal was amplified (Stanford Research Systems, SR570) and read by two lock-in amplifiers (Zurich Instruments MFLI) simultaneously, locked to frequencies $p$ and $f$ as $I_p$ and $I_f$, respectively. The data stream was recorded with buffer for 0.1 s to increase the sampling rate and fetched consecutively three times to improve the signal-to-noise ratio of each measurement. The out-of-plane magnetization of twisted MoTe$_2$ is proportional to the reflection difference between PEM-modulated right-handed and left-handed circularly polarized light, i.e. $\Delta R = I_f$. To account for the intensity fluctuation of laser excitation over time, the RMCD signal was normalized by the total reflection $R$, i.e. RMCD = $\Delta R / R = I_f / I_p$ x 100%.

Each probe pulse lasts for ~1 ms and multiple acquisitions within the probing window avoid the heating effect and ensure probing precision. In addition, the fast (50 kHz) and continuous variation of the helicity of probe light (i.e., one period includes linear → σ- → linear → σ+ → linear) minimizes the angular momentum transfer of probe to samples. Therefore, our adopted 6 nW·µm$^{-2}$ power of probe light does not interfere with the magnetic state of samples, as demonstrated in the **Extended Data Figure 8**.

For PL measurements, a 632.8 nm He-Ne linearly polarized laser passed through a narrow filter (Semrock LL01-633E-12.5) and was focused on the sample. The excitation power was adjusted using a neutral density filter and kept at 95 nW·µm$^{-2}$ for all measurements. The PL signal was collected by the same objective under a confocal setup after passing through a 980 nm long-pass filter (Semrock BLP01-980R-25) to remove the reflected laser scattering. Then, the PL photons were counted by a superconducting nanowire single photon detector (SNSPD, Single Quantum), or dispersed with a diffraction grating (300 grooves per mm at 1.2-µm blaze) before being detected by a liquid-nitrogen-cooled InGaAs charge-coupled device (CCD, Princeton Instruments PyLoN-IR 1.7).

**Calculation of Carrier Density and Electric Field**

The charge carrier density $n$ and electric field $D/\varepsilon_0$ in the sample were calculated from the top-gate ($V_{tg}$) and bottom-gate voltages ($V_{bg}$) using a parallel plate capacitor model:

$$n = \frac{V_{tg}C_{tg} + V_{bg}C_{bg}}{e} - n_0 \qquad (1)$$

$$D/\varepsilon_0 = \frac{V_{tg}C_{tg} - V_{bg}C_{bg}}{2\varepsilon_0} - D_0/\varepsilon_0 \qquad (2)$$

where $C_{tg}$ and $C_{bg}$ are the top and bottom gate capacitances obtained from the relative permittivity $\varepsilon_{hBN}$ and the measured BN thicknesses $d_{tg/bg}$ ($C_{tg/bg} = \frac{\varepsilon_{hBN}}{d_{tg/bg}}$), $e$ is the elementary charge, $\varepsilon_0$ is the vacuum permittivity. The adopted BN thicknesses were measured by the non-contact mode of an AFM (Park System NX10) with a precision of about 0.2 nm, which is negligible when considering the commonly used 15-30 nm thickness of BN. The most significant uncertainty comes from the relative permittivity of BN, which is reported in the range of 3-4 and has device-to-device variation due to crystal defects and dielectric interface quality. In this work, a value of 3.76 was adopted for the calculation according to the reference [62]. Both offsets of carrier density $n_0$, derived from the assignment of $v = \pm 1$ state filling factors in the PL measurement, and electric field $D_0/\varepsilon_0$, derived from the symmetric axis of the dual-gate RMCD map, were corrected. The carrier density at integer fillings calculated from the capacitor model was used to estimate the twist angle, yielding 3.50° for device #1 and 3.65° for device #2, which are consistent with the targeted twist angles.

**Determination of Topological Invariant**

The optically detected Streda slope in **Extended Data Figure 7** was extracted by tracking the evolution of the geometric centroid of PL central energy dips as a function of the applied out-of-plane magnetic field. The slope in the Streda formula represents the topological invariant according to $C = \Phi_0 \frac{\partial n}{\partial B}$, where $\Phi_0$ is the magnetic flux quantum, $n$ is the carrier density of the gapped state and $B$ is the magnetic field strength.

**Data Availability**

The data that support the findings of this study are available from the corresponding authors upon reasonable request.

23  Mangin, S. *et al.* Engineered materials for all-optical helicity-dependent magnetic switching. *Nature Materials* **13**, 286-292 (2014).
24  Siegrist, F. *et al.* Light-wave dynamic control of magnetism. *Nature* **571**, 240-244 (2019).
25  Zhang, P. *et al.* All-optical switching of magnetization in atomically thin $CrI_3$. *Nature materials* **21**, 1373-1378 (2022).
26  Xie, T. *et al.* High-efficiency optical training of itinerant two-dimensional magnets. *Nature Physics*, 1-7 (2025).
27  Chen, G. *et al.* Tunable correlated Chern insulator and ferromagnetism in a moiré superlattice. *Nature* **579**, 56-61 (2020).
28  Li, T. *et al.* Quantum anomalous Hall effect from intertwined moiré bands. *Nature* **600**, 641-646 (2021).
29  Anderson, E. *et al.* Programming correlated magnetic states with gate-controlled moiré geometry. *Science* **381**, 325-330 (2023).
30  Cai, J. *et al.* Signatures of fractional quantum anomalous Hall states in twisted $MoTe_2$. *Nature* **622**, 63-68 (2023).
31  Park, H. *et al.* Observation of fractionally quantized anomalous Hall effect. *Nature* **622**, 74-79 (2023).
32  Xu, F. *et al.* Observation of integer and fractional quantum anomalous Hall effects in twisted bilayer $MoTe_2$. *Physical Review X* **13**, 031037 (2023).
33  Zeng, Y. *et al.* Thermodynamic evidence of fractional Chern insulator in moiré $MoTe_2$. *Nature* **622**, 69-73 (2023).
34  Lu, Z. *et al.* Fractional quantum anomalous Hall effect in multilayer graphene. *Nature* **626**, 759-764 (2024).
35  Wu, F., Lovorn, T., Tutuc, E., Martin, I. & MacDonald, A. Topological insulators in twisted transition metal dichalcogenide homobilayers. *Physical Review Letters* **122**, 086402 (2019).
36  Li, H., Kumar, U., Sun, K. & Lin, S.-Z. Spontaneous fractional Chern insulators in transition metal dichalcogenide moiré superlattices. *Physical Review Research* **3**, L032070 (2021).
37  Tschirhart, C. *et al.* Intrinsic spin Hall torque in a moiré Chern magnet. *Nature Physics* **19**, 807-813 (2023).
38  Pershoguba, S. S. & Yakovenko, V. M. Optical control of topological memory based on orbital magnetization. *Physical Review B* **105**, 064423 (2022).
39  Adak, P. C., Sinha, S., Agarwal, A. & Deshmukh, M. M. Tunable moiré materials for probing Berry physics and topology. *Nature Reviews Materials* **9**, 481-498 (2024).
40  Ju, L., MacDonald, A. H., Mak, K. F., Shan, J. & Xu, X. The fractional quantum anomalous Hall effect. *Nature Reviews Materials*, 1-5 (2024).
41  Wang, C. *et al.* Fractional Chern insulator in twisted bilayer $MoTe_2$. *Physical Review Letters* **132**, 036501 (2024).
42  Kezilebieke, S. *et al.* Topological superconductivity in a van der Waals heterostructure. *Nature* **588**, 424-428 (2020).
43  Xu, F. *et al.* Signatures of unconventional superconductivity near reentrant and fractional quantum anomalous Hall insulators. *arXiv preprint arXiv:2504.06972* (2025).
44  Nayak, C., Simon, S. H., Stern, A., Freedman, M. & Das Sarma, S. Non-Abelian anyons and topological quantum computation. *Reviews of Modern Physics* **80**, 1083-1159 (2008).
20

**Acknowledgements**

We sincerely thank Xiaoqin Li, Duncan Haldane, Yoshinori Tokura and Trithep Devakul for fruitful discussions, and Xiaodong Xu and Jiun-Haw Chu for flux-grown MoTe$_2$ crystals. This work is supported by ASTAR (M21K2c0116, M24M8b0004), Singapore National Research foundation (NRF-CRP22-2019-0004, NRF-CRP30-2023-0003, NRF-CRP31-0001, NRF2023-ITC004-001 and NRF-MSG-2023-0002) and Singapore Ministry of Education Tier 2 Grant (MOE-T2EP50221-0005, MOE-T2EP50222-0018). X.C. acknowledges the support from the NTU Presidential Postdoctoral Fellowship (023455-00001). K.W. and T.T. acknowledge support from the JSPS KAKENHI (Grant Numbers 21H05233 and 23H02052) and World Premier International Research Center Initiative (WPI), MEXT, Japan.

**Author Contributions**

W.G. and X.C. conceived the project. H.P. and X.C. fabricated the devices with the assistance of S.Y. X.C. and H.P. performed the optical measurements and data analyses with the help of Y.Z., R.H. and J.Z.P. B.Y., Y.W. and A.R. provided theoretical input. W.W., R.D. and Z.L. synthesized MoTe$_2$ crystals. K.W. and T.T. provided high-quality h-BN crystals. X.C. prepared the manuscript and materials. All authors participated in the results discussion and contributed to the improvement of manuscript.

CRediT Taxonomy:

Methodology: X.C., H.P., Y.W., A.R.

Resources: W.G., B.Y., Z.L., J.Z.P., K.W., T.T.

Funding acquisition: W.G., B.Y., Z.L., J.Z.P., K.W., T.T.

Investigation: X.C., H.P., Y.W., S.Y., W.W., R.D., Y.Z.

Data curation: X.C., H.P., Y.W.

Validation: X.C., H.P., Y.W., W.G., B.Y., Z.L., J.Z.P.

Formal analysis: X.C., H.P., Y.W., A.R.

Software: Y.W., B.Y.

Visualization: X.C., Y.W.

Supervision: W.G., B.Y., J.Z.P., R.H.

Writing--original draft: X.C., Y.W.

Writing--review & editing: X.C., Y.W., W.G., B.Y., A.R., J.Z.P.

Project administration: X.C., W.G., B.Y., J.Z.P.

## **Competing Interests**

The authors declare no competing interests.

## **Additional Information**

Correspondence and requests for materials should be addressed to B.Y. (yang.bo@ntu.edu.sg) and W.G. (wbgao@ntu.edu.sg).